\documentclass[twocolumn,showpacs]{revtex4}
\usepackage{amsmath}
\usepackage{graphicx}
\usepackage{epsfig}
\usepackage{color}
\usepackage{bm}

\begin{document}

\title{Non-parametric segmentation of non-stationary time series}
\author{S. Camargo$^1$}
\author{S. Duarte Queir\'os$^2$}
\author{C. Anteneodo$^{1,3}$}
\affiliation{$^1$Departamento de F\'{\i}sica, PUC-Rio, Rio de Janeiro, Brazil \\
$^2$Istituto dei Sistemi Complessi - CNR, Roma, Italy \\
$^3$National Institute of Science and Technology for Complex Systems, Rio de
Janeiro, Brazil }

\begin{abstract}
The non-stationary evolution of observable quantities in complex systems can frequently be described
as a juxtaposition of quasi-stationary spells.
Given that standard theoretical and data analysis approaches usually  rely on the assumption
of stationarity, it is important to detect in real time series intervals holding that property.
With that aim, we introduce a segmentation algorithm
based on a fully non-parametric approach. We illustrate its applicability through the analysis of
real time series presenting diverse degrees of non-stationarity, thus
showing that this segmentation procedure generalizes and allows to uncover features
unresolved by previous proposals based on the discrepancy of low order statistical moments only.
\end{abstract}


\maketitle

\section{Introduction}

Complex systems are seldom in equilibrium or even in stationary states; however, 
their evolution can in many cases be thought as being composed of spells of
quasi-stationarity  in which  time-varying pseudo-parameters can be considered unchanged.
Examples of such framework can be found in finance~\cite{financegenrefs},
biology~\cite{biogenrefs}, physics~\cite{bodenschatz,physiogenrefs}
and physiology~\cite{physiogenrefs,hensel}, just to mention a few areas.
By identifying stationary segments, one can apply  standard techniques, e.g.,
extracting stochastic equations (Kramers-Moyal coefficients) from data ~\cite{friedrich},
overcoming the difficulties of non-stationary treatments~\cite{hensel,nonsteady}.
As another application, a proper segmentation is important to assess the scenario 
of mixed statistics~\cite{beck} based on the idea of local equilibrium,
typically applied by considering a fixed characteristic scale (window length). 
In general, segmentation provides a useful portrait of the local statistical properties  
for modeling non-stationary systems.

In order to identify such quasi-stationary patches,
algorithms based on  standard statistical methodology have been
introduced. Explicitly, they lean on moving along the series a pointer to detect the
position that maximizes a given quantifier of the statistical discrepancy
between the segments on both sides of the pointer. Among others~\cite{others}, 
worth of mention are the algorithms based on the  Student's $t$-statistic
(used to test the significance of the null hypothesis
of equal means)~\cite{bernaola,fukuda} or  on the Jensen-Shannon
divergence in the case of symbolic sequences~\cite{carpena.ea:11}.  

Despite the interesting results provided by these
methods~\cite{carpena.ea:11}-\cite{grosse.ea:02},
limitations hampering the performance can be found in every of them.
On the one hand, in the statistical moments criteria there 
is the problem of boiling down the existence of non-stationarity
to the change of pre-established local quantities. For instance,
even if the time series presents fluctuations in the variance, the $t$-test may give us the
indication that the series is stationary. Although it could be improved through
the unequal variance $t$-test statistic or also through   an  $F$-test,
it will still rely on assumptions over the moments and on the validity  of the central limit theorem.
On the other  hand,  entropy based methods are more fitted to symbolic sequences, 
while information is lost if
discretizing a real valued series by means of thresholds.
Moreover, the segmentation stopping criteria can be deemed arbitrary.
Its proposed improvement by means of the Bayesian Information Criterion can be
disputed as well since such a criterion often favors minimalist modeling~\cite{entropy}.
With the aim of surmounting those difficulties,
we introduce a \textit{fully} non-parametric segmentation approach by
using the Kolmogorov-Smirnov (KS) statistic, $D_{KS}$, 
which measures the maximal distance between the cumulative
distributions of two samples, as estimate of the discrepancy between segments. 
Note that it allows to test whether two samples come 
from the same distribution with no need to specify which is the common distribution. 
 

\section{KS-segmentation algorithm}

Our algorithm (named KS-segmentation) works as follows. 
Given a segment of a time series, $\{x_{i},\;\;i_{1}\leq i\leq
i_{n}\}$, a sliding pointer, at $i=i_{p}$,  is moved in order to compare the
two fragments $S_{L}\equiv \{x_{i_{1}},\ldots ,x_{i_{p}}\}$ and $S_{R}\equiv
\{x_{i_{p}+1},\ldots ,x_{i_{n}}\}$. The position $i_{p}$ of
the pointer is moved  so
that the sizes of the two segments ($n_{L}=i_{p}-i_{1}+1$ and
$n_{R}=i_{n}-i_{p}$) are at least unitary. Then, one selects the position
$i_{max}$ that maximizes the Kolmogorov-Smirnov
(KS) statistic $D\equiv D_{KS}(1/n_{L}+1/n_{R})^{-1/2}$,
between the two patches $S_{L}$ and $S_{R}$.

Once found the position $i_{max}$ of the maximal distance $D$,
$D^{max}$, one checks the statistical significance (at a chosen significance level 
$\alpha =1-P_{0}$) of a potentially
relevant cut at $i_{max}$ by comparison with the result that would be
obtained was the sequence random~\cite{bernaola}.
The potential cut ticks the first stage if $D^{max}$
exceeds its critical value, $D^{max}_{crit}$, 
for the selected significance level (see Fig.~\ref{fig:critical} and 
further technical details in  App.~\ref{app}). 
Before final acceptance of the cut, one can still require  a
minimal size (number of points) $\ell _{0}$, namely, $i_{max}-i_{1}+1$,
$i_{n}-i_{max}\geq \ell _{0}$. The procedure is then recursively applied starting
from the full series $\{x_{i},\;\;1\leq i\leq N\}$, where $N$ is the total
number of data points, until no segmentable patches are left.
 The search for $D^{max}$ within a given segment $\{i_{1}, \ldots, i_{n}\}$ 
during the iterations, as well as  in the determination 
of the critical curves, is performed  for  $i_{1} \le i_p \le  i_{n}-1$. 
The outcome of this segmentation procedure when applied to paradigmatic 
non-stationary time series is hereafter presented. 


\section{Applications}

\begin{figure}[h!]
\includegraphics*[bb= 50 410 510 660, width=0.99\columnwidth,angle=0]{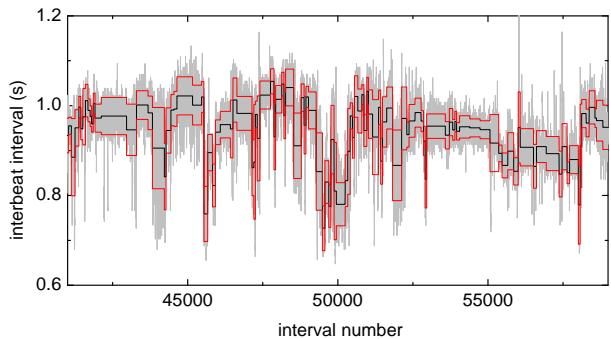}
\caption{(Color online) Fragment of a heart rate time series for a healthy individual  
(\texttt{n5nn.txt}) (light gray lines). The mean $+/-$ the standard deviation 
of the segments resulting from the KS-segmentation, with $\ell_0=50$  and  $P_{0}=0.95$,  
are displayed.
}
\label{fig:RR-series}
\end{figure}

We first survey the segmentation of heart-rate (RR) series, 
which motivated the introduction of the segmentation algorithm
based on the discrepancy of the means (mean-based algorithm)~\cite{bernaola} and 
that have also been suggested as a common focus to
solve the controversy over the potential chaoticity of normal heart rate~\cite{glass:09}.
Namely, we revisit the study of interbeat time series from healthy individuals 
(\texttt{nor}) and patients
with congestive heart failure (\texttt{chf})~\cite{physionet}.
Time series (tagged \texttt{n1nn-n5nn} and \texttt{c1nn-c5nn} for
(\texttt{nor}) and (\texttt{chf}), respectively)  are about 24 hours long and had their
outliers removed.
The segmentation outcome is depicted in Fig.~\ref{fig:RR-series}.  

\begin{figure}[ht!]
\includegraphics*[bb= 50 300 585 654, width=1.0\columnwidth,angle=0]{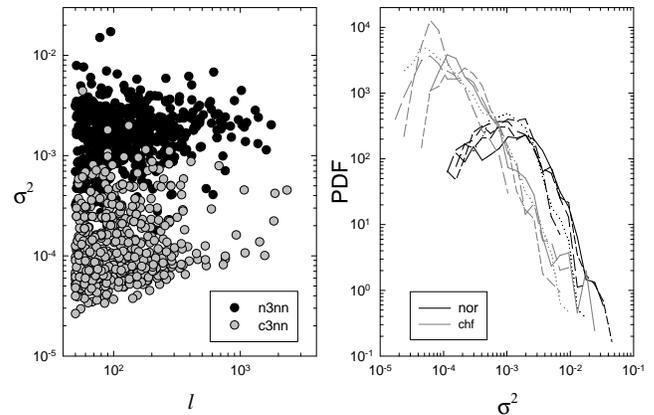}
\caption{ Local variance vs segment length for a representative individual of each
group (left).  Probability distribution function  (PDF) of the variance 
for all individuals of the \texttt{nor} (black lines) and
\texttt{chf} (gray lines) groups (right panel).}
\label{fig:dispersion}
\end{figure}

We computed the first moments for each resulting segment. 
The variance is deemed not constant throughout segments, but
it is dispersed over more than one decade as illustrated in Fig.~\ref{fig:dispersion},
for all segment sizes. The local variance is larger for the healthy subjects.
Thence, equal variance cannot be assumed as in previous
analysis of heart-rate series~\cite{bernaola}.
Furthermore, our finding implies that if one keeps such a simpler analysis,
at least the effective degrees of freedom in the $t$-test should be obtained
by the Welch-Sattherthwaite equation.  
It is worth referring that despite the tendency for the variance 
to be larger in the \texttt{nor} group, the
quotient for consecutive segments is very similarly distributed in both groups, 
with a slow power-law decay (with exponent close to -3) (not shown).

\begin{figure}[h!]
\includegraphics[scale=0.99]{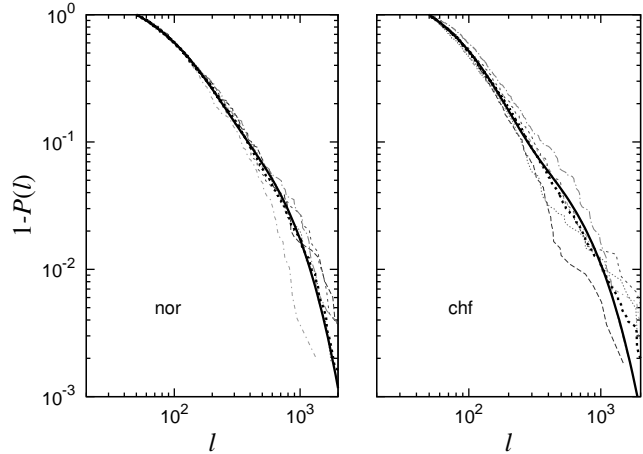}
\caption{Cumulative distribution of segment sizes for \texttt{nor} (left panel)
and  \texttt{chf} (right panel) individuals.
In each panel, the thin lines correspond to each individual of the group
(5 samples of 24-hour data), the dark dotted line to the entire group  and
the dark full line to  fits to
$A{\rm e}^{ -(l-\ell_0)/ L_1}+(1-A){\rm e}^{-(l-\ell_0)/L_2}$
with amplitude and characteristic lengths
$(A,L_1,L_2)=$  (0.78,78,372) and  (0.86,64,373) respectively.}
\label{fig:nor_chf}
\end{figure}

The complementary cumulative distributions of segment sizes for  \texttt{nor} and
 \texttt{chf}  individuals are displayed in Fig.~\ref{fig:nor_chf}.
For both groups, the plots can be described by a double exponential $A{\rm e}^{ 
-(l-\ell _{0})/L_{1}}+(1-A){\rm e}^{ -(l-\ell _{0})/L_{2} }$ with
characteristic lengths $L_{1}\simeq 70$ and $L_{2}\simeq 370$.
Hence there is no indication of a scale-free  behavior, as suggested by
previous segmentation analysis through the mean-based algorithm~\cite{bernaola}.
 
Our next example concerns the scenario of mixed statistics that has been applied to the study
of fluid turbulence~\cite{bodenschatz,beckturb}. Besides turbulence, its relevance is highlighted
by the fact that several models for finance have been inspired by this physical problem~\cite{beckvan}.
Succinctly, the mixed approach corresponds to a conjecture where one has a classical Boltzmann-Gibbs
statistics, conditioned to given temperature ($T\sim \beta^{-1}\sim \sigma^2$), 
which signals the existence of local equilibrium, that is associated with 
certain distribution $P\left( \beta \right)$.
Nonetheless, up to now, the approaches to the problem have considered the existence of a
\textit{single scale} of local equilibrium, which can be seen as a first
step for an outright description~\cite{beckvan,bcs}.
Endowed with our segmentation algorithm, we are in position
to evaluate the distribution of local time scales, $\tau$, of the $\beta$
factor and verify the local equilibrium assumption in this type of system.
As example, we consider a series of wind velocities 
(one month of measurements at a 30s acquisition interval)~\cite{wind}.
We reduced the strong daily periodicity by scaling the data by the average 
at each time of the day.
Then velocity $\vec{v}$ is dimensionless. 

\begin{figure}[b!]
\includegraphics*[bb= 80 480 510 750, width=0.99\columnwidth,angle=0]{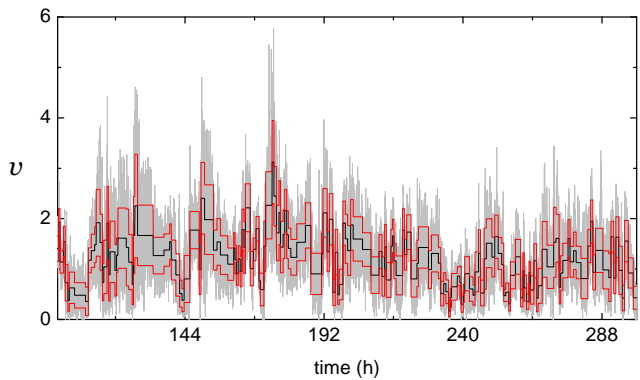} 
\caption{(Color online) Time series of   $v=|\vec{v}|$ (dimensionless).  
The mean $+/-$ the standard deviation of the segments resulting form the KS-segmentation
are displayed   (for $\ell_0=50$, $P_{0}=0.95$).
}
\label{fig:wind-series}
\end{figure}

\begin{figure}[h!]
\includegraphics*[bb= 70 400 520 700, width=0.90\columnwidth,angle=0]{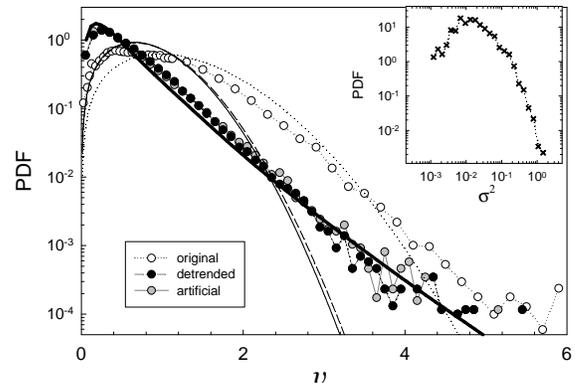}
\caption{ Distribution of  velocity  $v$ (circles).
Data were detrended by substracting the excess average with respect to
the one given by the Maxwell distribution with the same local variance
(black  circles). Distribution of an artificial series with the same segments as the
real one, with values of $v$ independently drawn from  2D-Maxwell distributions
with the local variance of the real series( gray circles).
Thin lines correspond to the respective 2D-Maxwell distributions, with the variance
of the whole series, and the full line to the mixing (numerically summed up) of 
the 2D-Maxwell with the distribution of the local variance obtained 
from the segmentation process, shown
in the inset.}
\label{fig:mixing}
\end{figure}

Accordingly, in a segment within which local equilibrium holds,
the velocity distribution is defined by
$p\left( \vec{v}| \beta  \right) =\frac{\beta }{2\pi }\exp
\left[-\beta \frac{|\vec{v}|^{2}}{2}\right] $, wherefrom the distribution of the
speed, $v\equiv |\vec{v}|$, is $p\left( v|\beta \right) =\beta \,v\,\exp
\left[ -\beta \frac{v^{2}}{2}\right] $.  Along these lines, we apply our
segmentation  procedure pitching at detecting the time intervals where the
local equilibrium approximation is valid. The result of the segmentation is
depicted in Fig.~\ref{fig:wind-series}. The complementary cumulative distribution of segments 
decays more slowly than exponentially (plausibly a stretched or double exponential,
the latter with characteristic times of 32 min and 93 min).
The distribution of segment lengths has mean (standard deviation) approximately equal to
129 (91) points  (corresponding to 64  (45) min approx.).
One observes in Fig.~\ref{fig:mixing} that the 2D-Maxwell distribution fails in
describing the distribution of velocities, because the local variance is dispersed
(inset of Fig.~\ref{fig:mixing}). Then we considered the mixing
$p(v)=\int_0^\infty d\sigma ^2 p \left(  v | \sigma^2  \right) p(\sigma^2)$,
where $ p(v | \sigma^2) $ is the Maxwell distribution defined above, 
substituting $\beta=\left( 4-\pi\right) /\left( 2\sigma ^{2} \right)$,
given that the  (conditioned) raw moments
are $\left\langle v^{n}\right\rangle_{\beta }=\left( 2/\beta \right)^{n/2}\,
\Gamma \left[ 1+n/2\right]$.
One observes that the mixed distribution is in good accord with the data distribution,
once local trends are removed. The mixed distribution also agrees with the histogram 
built from artificial series obtained by juxtaposition of sequences of 
Maxwellian random numbers, with the same length and local variance as the real ones.

Afterwards, we have computed the local variance
that is proportional to the inverse $\beta$ factor.
Its  distribution,  presented in the inset of
Fig.~\ref{fig:mixing},  is responsible for the deviation of
$p\left( v\right)$ from the 2D-Maxwell distribution (main frame of Fig.~\ref{fig:mixing}).

\begin{figure}[!h]
\includegraphics*[bb= 70 490 500 750, width=0.99\columnwidth,angle=0]{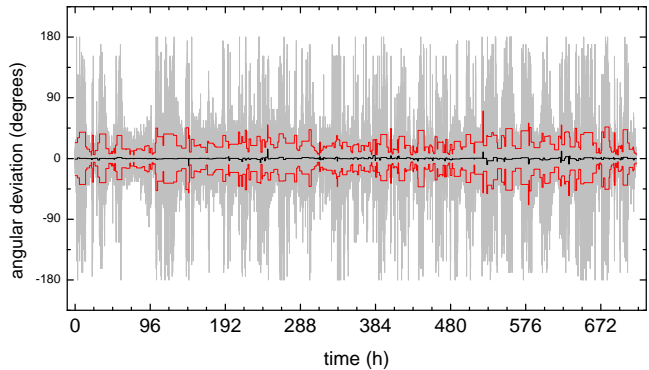}
\caption{(Color online) Segmentation of angular deviation at 30s lag. The mean $+/-$ 
the standard deviation of the segments resulting
from the KS-segmentation are displayed.}
\label{fig:series-angle}
\end{figure}

Besides, the speed, $v$, we also looked at angle variations (at 30s lag). 
While a more quantitative analysis on this matter is addressed to future work, 
here we would like to call attention to the fact that, 
even if the average value is constantly close to zero, 
changes in the local variance are detected by the present method, as 
depicted in Fig.~\ref{fig:series-angle}, while no segmentation occurs with the 
procedure based on the discrepancy of the means. 

\section{Final remarks}

In this manuscript we have presented a segmentation method which aims 
at coping with non-stationary signals from widespread 
physical and non-physical systems where the local-equilibrium
or local-stationarity hypotheses hold. Our method, 
which is based on the Kolmogorov-Smirnov test, 
improves previous proposals as soon as it is non-parametric and 
thus independent of any pre-assumed 
order of fluctuation between the stationary segments,  resulting into 
a more flexible and effectual algorithm. 

Concerning algorithmic complexity, our algorithm is more efficient 
than methods based on  matrix diagonalization, such as principal component analysis. 
In comparison with moment based segmentation proposals, our algorithm requires  sorting 
 segments of length $n$ which increases the complexity in a factor of order $\ln n$, 
 which does not represent a significant larger computational cost, despite the enhanced  
 ability.

The applicability of our proposal was then tested with two poles apart signals, 
heart-rate intervals and wind  velocities, 
with significant results in each case. 
In the first case, the non-stationarity portrait   is altered with respect to that of previous 
analysis based on the discrepancy of the means, as soon as the local variance 
cannot be assumed constant. 
In the second, the procedure is shown to be useful to detect meaningful windows to compute 
local statistical quantities. 
In general, proper segmentation is helpful in several problems where 
local-stationarity  applies.

\appendix
\section{}
\label{app}

\subsection{Statistical significance criterion}

\begin{figure}[h!]
\includegraphics*[bb= 40 440 510 750, width=0.95\columnwidth,angle=0] {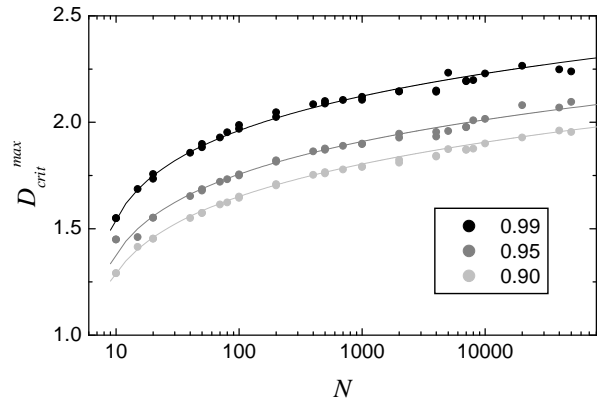}
\caption{Critical values of $D^{max}$ as a function of the sequence
length $N$ for series of Gaussian i.i.d. random numbers, 
at different significance levels $P_{0}$ indicated on
the figure. Full lines are fits to the data,  used as phenomenological
formulae for significance checking.}
\label{fig:critical}
\end{figure}

We determined $D^{max}$ numerically for a large number ($>10^{4}$) of sequences of $N$ i.i.d.
Gaussian numbers and built its cumulative distribution. 
From the cumulative distribution, we obtained the critical values of $D^{max}(N)$, 
$D^{max}_{crit}(N)$, for each given significance level $\alpha =1-P_{0}$. The resulting 
critical curves are  shown in Figure~\ref{fig:critical}. 
For the significance tests applied throughout the segmentation procedure, 
we used the effective form of 
the critical curves given by the heuristic simple expression  
\begin{equation}
D^{max}_{crit}(N)=a(\ln N-b)^{c},
\end{equation}
with $(a,b,c)$ equal to $(1.41,1.74,0.15)$, $(1.52,1.8,0.14)$ and $(1.72,1.86,0.13)$ 
for $P_{0}=0.90$, $0.95$ and $0.99$, 
respectively.

We noticed that, along a random series, the position $i_{\max}$ for which $D$ 
is maximal is not uniform but presents a U-shaped distribution. 
This could set forth a bias in the cutting performance propping up an increase 
in the number of short segments. 
Let us mention that in the mean-based algorithm an alike U-shape is also present.
To avoid this effect, we tested a redefinition of the standard KS distance, by 
considering $D \equiv D_{KS}(1/n_{1}+1/n_{2})^{- \gamma}$, with arbitrary $\gamma$ 
and observed that the flattest (more uniform) distribution of $i_{max}$, for any 
size $N$, occurs for $\gamma\simeq 0.64$.  
We compared the implementation of the algorithm both values of $\gamma$ (0.5 and 0.64), 
however, both for real and artificial series no significant differences in the 
segmentation portrait were observed. Then we kept the standard definition of $D$.

\subsection{Testing artificial series}

\begin{figure}[h!]
\centering
 \includegraphics*[bb=70 430 530 760, width=0.5\textwidth]{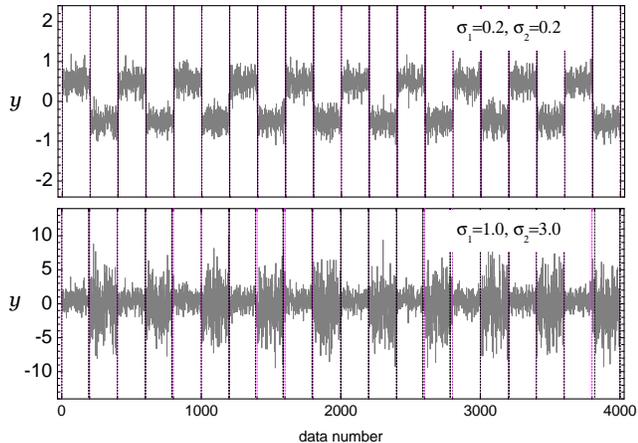}
\caption{Artificial series (lightgray lines) formed by segments of  $n=200$ 
Gaussian numbers with alternating 
means $+1/2$, $-1/2$ and standard deviation $\sigma_1$, $\sigma_2$ 
(values indicated on each panel). 
Segmentation was performed by means of the KS-algorithm at level $P_0=0.95$. 
The vertical dotted lines indicate the exact (magenta) and calculated (black) borders. 
}
\label{fig:artificial}
\end{figure}

To check the performance of the algorithm, we analyzed artificial series 
$\{y_i, 1\le i\le N\}$, 
formed by segments of  $n$ Gaussian numbers. We set unitary jumps in the means of 
consecutive segments ($\Delta \bar{y}=1$) and alternating 
standard deviation (square root of the variance) $\sigma_1$, $\sigma_2$,   as illustrated in Fig.~\ref{fig:artificial}.  
We varied each standard deviation from 1/10 to 10, then embracing a wide range of values relative to 
the jump size.
 
\begin{figure}[h!]
\centering
 \includegraphics[scale=1.1]{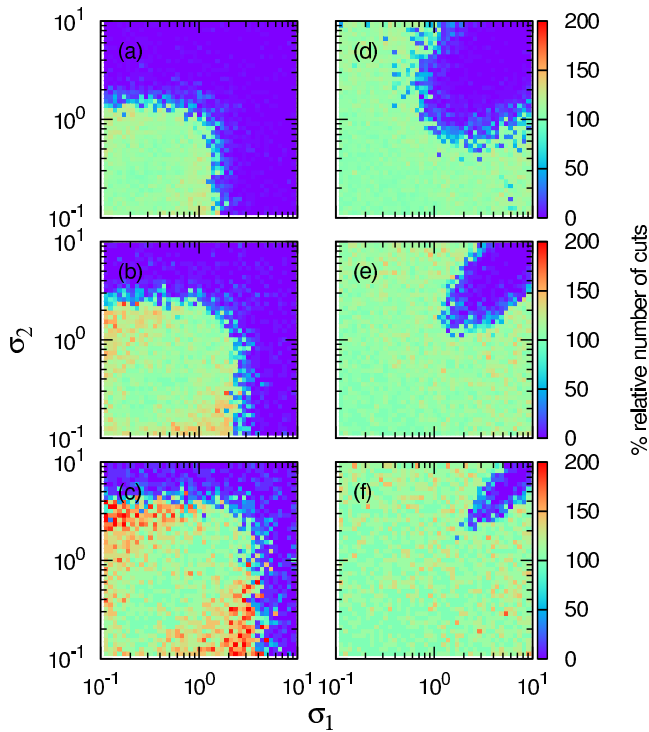}
\caption{(Color online) Segmentation diagram in the parameter plane $\sigma_1,\sigma_2$. 
The percentual relative number of cuts is represented in a palette mapping. 
Each grid cell corresponds to a different random sequence of size $N=4000$ and 
segment sizes $n=100$ (a,d), 200 (b,e) and 400 (c,f). 
Segmentation was performed with the mean-based (a-c) and KS (d-f) algorithms, 
with $\ell_0=10$ and $P_0=0.95$. 
}
\label{fig:diagram10}
\end{figure}

\begin{figure}[h!]
\centering
 \includegraphics[scale=1.1]{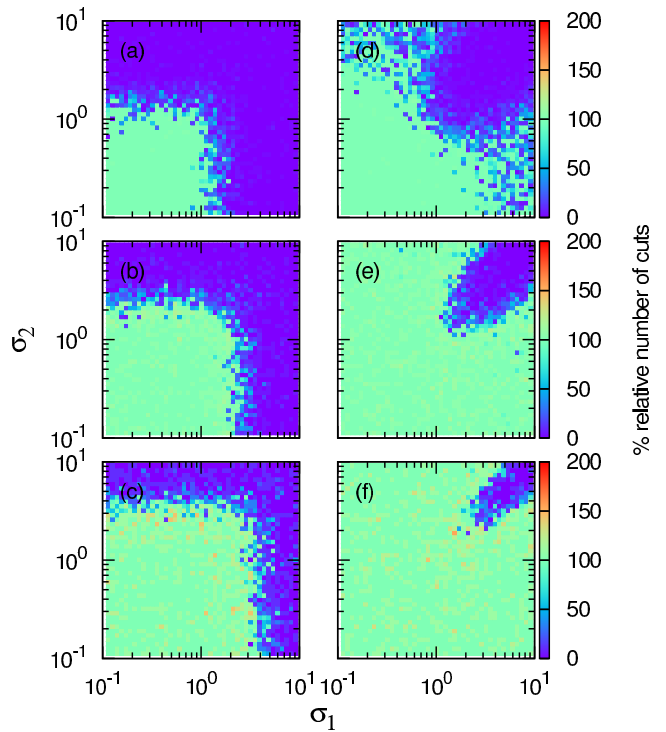}
\caption{(Color online) Segmentation diagram in the parameter plane $\sigma_1,\sigma_2$, 
as in Fig.~\ref{fig:diagram10} but with with $\ell_0=50$. 
}
\label{fig:diagram50}
\end{figure}

Diagrams of the segmentation results in the plane $\sigma_1$, 
$\sigma_2$ are shown in Figs.~\ref{fig:diagram10} and \ref{fig:diagram50}, for $\ell_0=10$ and 50,  
respectively, at level $P_0=0.95$. For each sequence, the percent relative number of cuts 
with respect to the actual one is displayed. 
The outcomes of the KS-algorithm are shown in the right-hand side panels and 
those of the mean-based algorithm  are also presented (left-hand side panels) for comparison. 
Time series belonging to the cyan (light) regions are 
correctly (100\%) segmented, while those belonging blue (dark) regions are typically unsegmentable. 
Red cells indicate oversegmentation. 

\begin{figure}[h!]
\centering
 \includegraphics[scale=1.1]{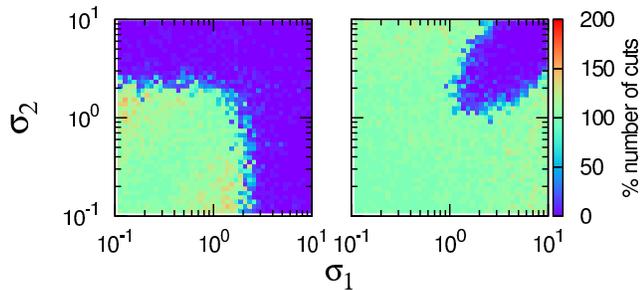}
\caption{(Color online) Segmentation diagram in the parameter plane $\sigma_1,\sigma_2$, 
as in Fig.~\ref{fig:diagram10}(b,e) (i.e., $n=200$, $\ell=10$) but with with $P_0=0.99$. 
}
\label{fig:diagram099}
\end{figure}

The mean-based algorithm performs proper segmentation only if the standard deviations 
are at most of the order of the 
size of the jumps and works well, within the chosen confidence level, 
around de diagonal ($\sigma_1\simeq \sigma_2$), as expected. 
Meanwhile, the KS algorithm is able to segment series in a larger region of parameter space. 
Segmentation fails when the standard deviations of consecutive segments are not significantly 
different and are larger than the jump size ($\sigma_1\simeq\sigma_2>\Delta \bar{y}$).
In both procedures, for larger segment sizes $n$, the segmentable domain enlarges, but more 
false cutting points arise. By setting a larger value of $\ell_0$, small segments 
are discarded and the number of false cuts is reduced, as can be seen by 
comparison of Figs.~\ref{fig:diagram10} and \ref{fig:diagram50} which just differ in the value 
of $\ell_0$.
Of course, false  cutting points can also be reduced by increasing the value of $P_0$ (compare 
the second row of Fig.~\ref{fig:diagram50} with Fig.~\ref{fig:diagram099}, that only  
differ in the value of $P_0$.

\begin{figure}[h!]
\centering
 \includegraphics[scale=1.1]{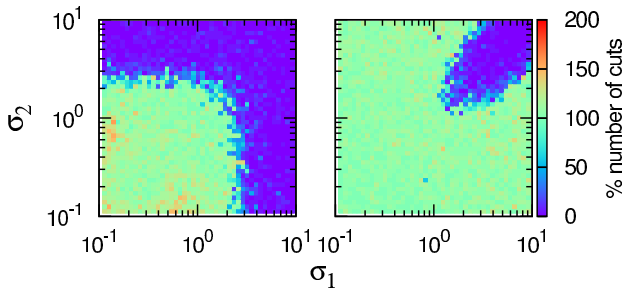}
\caption{(Color online) Segmentation diagram in the parameter plane $\sigma_1,\sigma_2$, 
as in Fig.~\ref{fig:diagram10}(b,e) (i.e., $n=200$, $\ell=10$, $P_0=0.95$), 
with the additional neighbor segment check. 
}
\label{fig:diagramNN}
\end{figure}

Moreover, one could still improve the algorithm by adding a further 
final step, which is the following one. Before accepting a cut, check  through the standard  
KS test the significance of the discrepancy between the right-hand side portion and its right neighboring 
segment (born in the previous generation) as well as the left-hand side portion with its left neighbor, 
as it has been proposed for the mean-based segmentation~\cite{bernaola}. 
For the analyzed series, this step does not introduce a significant improvement 
(see Fig.~\ref{fig:diagramNN} to be compared with the second row of Fig.~\ref{fig:diagram10}), 
however, if,  depending on the analyzed series, one observes oversegmentation, 
then that step could be straightforwardly added.   

Let us comment that as one approaches the frontier of the segmentable region, 
although segments are recognized, 
the position of the boundaries of the segments gets more imprecise. 
Effect which is reduced by increasing the confidence level. 

The above diagrams depict the scope of the algorithm 
for a particular class of artificial series but provides a feeling of 
its range of applicability  and limitations. 
It also manifests the importance of our method in enlarging the domain of segmentation.
There is an infinity of other tests, e.g., with diverse variabilities of means and 
variances, correlations, other statistics discrepancies, that could be performed. 
Also tests restricted to the comparison of the outcoming statistics could 
be carried out~\cite{fukuda}. 
However, when applying this or other algorithm, it may be convenient to perform ad-hoc 
test, depending on the particular statistical characteristics of the analyzed series.

\subsection{Robustness}

\begin{figure}[h!]
\includegraphics[scale=0.95]{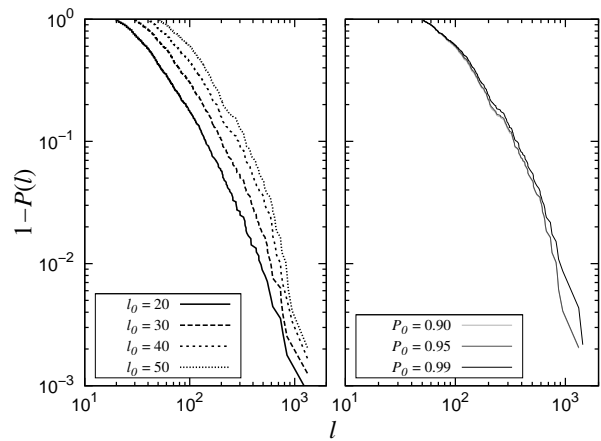}
\caption{Left panel: Complementary cumulative distribution of segment sizes 
obtained for different values of 
the minimal length $\ell_0$ indicated on the figure. 
Data correspond to a normal individual (\texttt{n5nn.txt}). 
Right panel: Complementary cumulative distribution of segment sizes 
at different significance levels indicated on the figure.}
\label{fig:robustness}
\end{figure}

For the analyzed series, we checked the robustness of the results with respect to 
the significance level ($P_0=0.90,0.95,0.99$). 
The smaller $P_0$, the larger the tendency to allow small segments, while
larger ones are almost unaffected. However, this effect does not change 
significantly the statistics of segment sizes, as illustrated 
for heart-rate series in  Fig.~\ref{fig:robustness} (right panel). 
The impact of $\ell_0$ was also checked (left panel of  Fig.~\ref{fig:robustness}). 
Slopes do not significantly change, except for
small values of $l$, as expected, since smaller fragments are allowed with
decreasing $\ell_0$. Notwithstanding, the probability density of larger segments 
is not significantly altered. 
Let us notice that the statistics on segments may be
affected by the increase of small segments if the studied quantity is correlated with
the size. However, this does not seem to happen in the analyzed cases. 

We also checked that the same segmentation patches are typically  recovered 
even when analyzing small fragments 
of the whole series. In fact, significant statistical jumps 
even for small segments are recognized, prompting a segmentation point.


\section*{Acknowledgements:}
We are grateful to Pedro Bernaola-Galv\'an for useful exchange of correspondence 
about previous segmentation proposals. Also acknowledged are Salvo Rizzo and 
Stefano Ruffo for  aiding us to obtain wind data.
We acknowledge Brazilian agencies Faperj (Foundation for Research Support,
State of Rio de Janeiro) and CNPq (National Council for Scientific and
Technological Development) and the European Commission through the 
Marie Curie Actions FP7-PEOPLE-2009-IEF (contract nr 250589) for financial support.

\end{document}